\documentclass[10pt,a4paper]{article}
\usepackage[toc,page]{appendix}
\usepackage{fullpage}
\usepackage[latin1]{inputenc}
\usepackage[french,english] {babel}
\usepackage{amsmath}
\usepackage{amsfonts}
\usepackage{amssymb}
\usepackage{graphicx}
\usepackage{epstopdf}
\usepackage{color}
\usepackage{listings}
\usepackage[stable]{footmisc}
\usepackage{verbatim}
\usepackage{subfig}
\usepackage[top=2.5cm,bottom=2.5cm,right=2.5cm,left=2.5cm]{geometry}
\usepackage{amsthm}
\usepackage{feynmp}
\usepackage{array}
\definecolor{vert}{rgb}{0,0.5,0}
\newcommand{\beqn}{\begin{eqnarray}}
\newcommand{\eeqn}{\end{eqnarray}}
\newcommand{\beqas}{\begin{eqnarray*}}
\newcommand{\eeqas}{\end{eqnarray*}}
\newcommand{\beq}{\begin{equation}}
\newcommand{\eeq}{\end{equation}}
\newcommand{\bit}{\begin{itemize}}
\newcommand{\eit}{\end{itemize}}
\newcommand{\bseq}{\begin{subequations}}
\newcommand{\bal}{\begin{align}}
\newcommand{\eal}{\end{align}}
\newcommand{\eseq}{\end{subequations}}
\newcommand{\nn}{\nonumber}
\newcommand{\bs}{\boldsymbol}

\begin{document}
\begin{flushright}
{ULB-TH/14-06}  \\
{INR-TH-2014-005}
\end{flushright}
\vskip 1cm
\begin{center}
{\huge CP violation from pure gauge in extra dimensions}
\vskip .7cm
{\large Jean-Marie Fr\`ere${}^a$},
{\large Maxim Libanov${}^{b,c}$} and
{\large Simon Mollet${}^a$}
\vskip .7cm
\emph{${}^a$Service de Physique Th\'eorique,\\
Universit\'e Libre de Bruxelles, ULB, Campus de la Plaine CP225, Bd du Triomphe, 1050 Brussels, Belgium
\vskip 0.2cm
${}^b$Institute for Nuclear Research of the Russian Academy of Sciences,\\
60th October Anniversary Prospect 7a, 117312, Moscow, Russia
\vskip 0.2cm
${}^{c}$  Moscow Institute of Physics and Technology,\\
Institutskii per., 9, 141700, Dolgoprudny, Moscow Region, Russia}
\end{center}

\begin{abstract}

One of the Sakharov's condition for baryogenesis is the violation of both C and CP. In the Standard Model, gauge interactions break maximally C, but CP is only broken through the Yukawa couplings in the poorly understood scalar sector. In extra-dimensional models, extra components of gauge fields behave as scalars in 4D and can acquire effective vev's through (finite) quantum effects (Hosotani mechanism). This mechanism is used to build a \textit{toy model} with 2 extra-dimensions compactified on a flat torus $T^2$, where a SU(2) gauge symmetry is broken to U(1) and CP violation (in 4D) is expected. This is verified by computing a non-vanishing electric dipole moment.

\end{abstract}

\bigskip

\section{Introduction}

In comparison with "pure gauge" theories, scalar interactions are badly
understood --- our ignorance being parametrized  through a bunch of
arbitrary (Yukawa) couplings. Moreover, while the gauge interactions are
CP conserving (at least in 4D)\footnote{We ignore mass terms which are, at
least in chiral theories, a counterpart of scalar interactions.}, the
scalars  break this symmetry, but still in an arbitrary manner (through
the phases of the Yukawa coefficients). The situation is well-known in the
Standard Model (SM) where CP-violating freedom is only empirically
constrained. It is then a sensible belief that a better insight in the
scalar sector could clarify the nature of CP violation, and vice versa.

Possibly more central than CP symmetry itself is the issue of
matter-antimatter asymmetry. As was pointed out by Sakharov, the emergence
of a matter-antimatter asymmetry from an initially symmetrical early
universe requires in 4D both C and CP violation. A more general statement
would be that C and any symmetry involving C must be broken, CP being just
one particular case. This is pretty much the situation we will be
discussing in the present note: how C or CP invariance can be broken in
theories containing only fermions and their gauge interactions. More
specifically, we will discuss how C or CP conservation behave in the
dimensional reduction (in the present case from 6D to 4D).

In an attractive, though quite old idea, scalar fields are thought as
spatial components of gauge fields in extra dimensions (ED)
\cite{Hosotani:1983a}\cite{Hosotani:1983b}\cite{Hosotani:1989}. When
extra-dimensional space is not simply connected, non trivial holonomies
(or Wilson lines (WL)) can appear dynamically for non contractible
cycles\footnote{This can be seen (at least for abelian cases) as finite
magnetic fluxes through holes in the manifold. However, these holes being
outside the physical space, a flux is always ill-defined, hence the use of
holonomies.} and lead to dynamical symmetry breaking. At the level of our
(3+1)-dimensional space, effective scalar fields acquire a vev, which
could cause CP violation if scalar and pseudo-scalar contributions
coexist. At the classical level, the WL are determined by the topology of
ED and label degenerate classical vacua. The degeneracy disappears when
quantum effects are taken into account, which select the physical
solution. These are encoded into the effective potential for WL which
depends on topology, matter content and Scherk-Schwarz (SchSch) phases
(see below).

In a previous work, this idea was already used and revealed to be
promising \cite{Cosme:2003}\cite{Cosme:2004}. One extra dimension was
introduced, and the 5th components of gauge fields can yield the
equivalent of pseudoscalar terms in the 4D-reduced Lagrangian, leading to
a complex mass matrix and possible CP violation. Of course, this is not
enough, since we can always use a chiral rotation to make them real.
Therefore real masses (or in other words, half of the scalar sector) were
put in by hand\footnote{We will return to this question later; in
particular if complex mass terms are needed to generate CP \`{a} la
Kobayashi-Maskawa, other sources of CP violation (through the Kaluza-Klein
excitations for instance) remain in principle possible.}. An appealing
extension would be to add a second ED which will provide for this. This is
in some way the situation we will be dealing for.

Before turning to 6D however, we should stress that this previous work
viewed the Hosotani loops purely as external boundary conditions rather
than dynamical variables (in the way of the Bohm-Aharonov effect). Here we
will follow Hosotani's view, which sees these loops as dynamical
variables, and requires the evaluation of the effective Lagrangian, beyond
the tree level.

The problem proves difficult, and the present note deals with "proof of
concept", namely the possibility of CP violation in 4D from pure gauge
theory in 6D, but does not propose a realistic model. This is notably due
to the difficulty of generating a "low mass scale", providing non-zero
mass to the zero modes of the compactified theory: in the present note, we
will deal either with a massless low-energy sector separated from the
Kaluza-Klein scale, or accept small masses controlled by arbitrary phases
in the boundary conditions.

The paper is organized as follows. In section \ref{sec:PCCP} we review the
notions of P, C and CP symmetries in 4D and in 6D and link them through
compactification schemes. Section \ref{sec:Hosotani} is devoted to
Hosotani mechanism which takes place when compactification implies non
simply connected ED. We summarize it in the special case of the flat torus
$T^2$ and try to include Hosotani's approach in the more modern one
\cite{Alfaro:2007}\cite{Salvatori:2007}. In section \ref{sec:CPviolbyCB}
we use explicitly Hosotani mechanism to break CP through compactification
and give simple examples in section \ref{sec:Examples}. Finally, in
section (\ref{sec:ChAno6D}) we come back on anomaly issues which appear in
chiral theories that we have neglected before. Conclusions and
perspectives can be found in section \ref{sec:Conclusion}.

\section{P, C and CP in 4 and 6 dimensions}\label{sec:PCCP}

We use the notation $\gamma^{\mu}$ (resp. $\Gamma^A$) for 4D (resp. 6D)
gamma matrices\footnote{Our choice of representation can be found in
Appendix \ref{AppA1}.}. The parity transformation is given respectively by
$\mathcal{P}^{-1} \psi(t,\bs{x})\mathcal{P}=\gamma^0 \psi(t,-\bs{x})$ in
4D and $\mathcal{P}^{-1} \Psi(t,\bs{x})\mathcal{P}=\Gamma^0
\Psi(t,-\bs{x})$ in 6D, where $\psi$ and $\Psi$ are 4- and 6-dimensional
Dirac spinors \cite{Gavela:1984}. Charge conjugation is given by
$\mathcal{C}^{-1}\psi(x)\mathcal{C}^{-1}=C^{(4)}\gamma^0 \psi^{*}(x)$ and
$\mathcal{C}^{-1}\Psi(x)\mathcal{C}^{-1}=C^{(6)}\Gamma^0 \psi^{*}(x)$
where $C^{(4)}$ (resp. $C^{(6)}$) is a matrix which satisfies
${C^{(4)}}^{-1}\gamma^{\mu}C^{(4)}=\pm \gamma^0
{\gamma^{\mu}}^{*}\gamma^0$ (resp. ${C^{(6)}}^{-1}\Gamma^{A}C^{(6)}=\pm
\Gamma^0 {\Gamma^{A}}^{*}\Gamma^0$). The $+$ sign in these relations can
be used only in the absence of mass term (which is our case) and there is
then an ambiguity in the definition, but we will see that this is
unimportant for our purpose.

In 4 dimensions the two solutions are $C^{(4)}_1=\gamma_0\gamma_2$ and
$C^{(4)}_2=\gamma_1\gamma_3$ (up to phase factors), while in 6 dimensions
we find $C^{(6)}_1=\Gamma^0\Gamma^2\Gamma^4$ and
$C^{(6)}_2=\Gamma^1\Gamma^3\Gamma^5$. In even dimensions, the spinors can
be decomposed in two semi-spinors (or Weyl spinor) with the help of the
chirality projectors\footnote{In 4D the $+$ sign is identified with $L$
and the $-$ sign with $R$.} $P_{L/R}=\frac{1\pm\gamma^5}{2}$ (resp.
$P_{\pm}=\frac{1\pm\Gamma^7}{2}$). Since $\gamma^5$ anticommutes with all
$\gamma$'s (as well as does $\Gamma^7$ with all $\Gamma$'s) it is obvious
that charge conjugation in 4D links $\psi_L$ and $\psi^*_R$ (and vice
versa), while in 6D it links $\Psi_+$ with $\Psi^*_+$\footnote{This is
related to the fact that in 4D (resp. 6D) $\psi_L$ and $\psi^*_R$ (resp.
$\Psi_+$ and $\Psi^*_+$) are equivalent representations of the Lorentz
group.}. On the contrary, the parity connects $+$ and $-$ spinors in all
cases ($L$ and $R$ in 4D). Then the CP operation which is the combination
of these two connects $L$ and $L$ spinors in 4D, but $+$ and $-$ in 6D. As
announced this is completely independent of the choice for $C^{(4)}$
(resp. $C^{(6)}$).

Now what does it mean? Since gauge interactions connect spinors of the
same chirality, gauge symmetries give no reason  to introduce both
chiralities on an equal footing. Then, in all generality, P is not an
automatic symmetry of gauge interactions in both 4 and 6 dimensions.
However, while C symmetry is not automatic in 4D, this is always the case
in 6D, and conversely for CP. For this reason we need scalar interactions
in 4 dimensions to break CP (at perturbative level). In contrast if we
write a theory in 6 dimensions with only (say) a $+$ spinor then we break
CP. Does it mean that the resulting effective 4D theory is not CP
conserving? In other words, are the notions of CP in 4 and 6 dimensions
directly related to each other? The answer is no.

To realize this we need to find a relation between 4D and 6D CP
transformations. Let us focus on + spinor in 6D which is a Dirac spinor at
the 4D level (with $L$ and $R$ components). We know that C transforms
$\Psi_+(x)$ into $\Psi_+^c(x)\sim \gamma^5\gamma^2\Psi^*_+(x)$. On the
other hand, $+$ and $-$ components being representations of the rotation
group, we can use them to link $\Psi_+^c(x)$ with $\Psi_+^{CP_4}(x)\sim
\gamma^0 \gamma^2 \Psi^*_+(t,-x_1,-x_2,-x_3,x'_4,x'_5)$, where
$(x'_4,x'_5)$ result from a rotation of $(x_4,x_5)$. Indeed,
$\Psi_+^{CP_4}(x)$ is then a CP transformation at the 4D level. One
solution is to use a $\pi$-rotation in the $1-2$ and $3-5$ planes. Then
$(x'_4,x'_5)=(x_4,-x_5)$. But any additional rotation in the $4-5$ plane
leads to a valid definition\footnote{Note that the effect of this rotation
on the 4D fermion is obviously a chiral rotation.}. Since this combination
of transformations is a symmetry of the 6D theory, the 4D effective theory
will be CP violating only if the compactification is incompatible with all
the symmetries: \beqn \left\{
\begin{aligned}
& \Psi_+ \rightarrow \Psi^*_+\\
& X\equiv(x_4,x_5)^T \rightarrow \mathcal{R}\sigma^3
X=\mathcal{R}(x_4,-x_5)\equiv \hat{X}=(\hat{x}_4,\hat{x}_5).\\
\end{aligned}
\right.
\label{CP4d6d}
\eeqn for any rotation $\mathcal{R}
$. In other words, the 4D theory will be CP violating if we fail to find a
chiral rotation which reabsorbs the phases.

Let us take a simple example to illustrate this. Consider a flat torus
$T^2$ of radii $R_4=R_5=R$ with the following SchSch boundary conditions
(BC)\footnote{For now on $\Psi$ means $\Psi_+$ unless otherwise stated.}:
$\Psi(x_4+2\pi R,x_5)=e^{i\beta_1}\Psi(x_4,x_5)$ and $\Psi(x_4,x_5+2\pi
R)=e^{i\beta_2}\Psi(x_4,x_5)$. Under the prescribed transformation these
BC become $\Psi(\hat{x}_4+2\pi R\cos\theta,\hat{x}_5+2\pi R
\sin\theta)=e^{-i\beta_1}\Psi(\hat{x}_4,\hat{x}_5)$ and
$\Psi(\hat{x}_4+2\pi R \sin \theta,\hat{x}_5-2\pi R
\cos\theta)=e^{-i\beta_2}\Psi(\hat{x}_4,\hat{x}_5)$. The first relation is
compatible only if $\theta=\pi$ or if $\theta=0$ and $\beta_1 \in
\left\lbrace 0,\pi \right\rbrace$, while the second one is compatible only
if $\theta=0$ or $\theta=\pi$ and $\beta_2 \in \left\lbrace 0,\pi
\right\rbrace$. Then BC break effective 4D CP symmetry as soon as
$\beta_1$ and $\beta_2$ are both different from $0$ and $\pi$. The result
is of course independent of $\theta$.

Note by the way that we can proceed in the same way for $P$ and $C$. It is
straightforward to show that P invariance requires compatibility with the
transformation $X \rightarrow \mathcal{R}\sigma^3 X$, while C requires
compatibility with $\Psi \rightarrow \Psi^*$ and $X\rightarrow
\mathcal{R}X$. In our previous example, P is broken but not C (this leads
then to CP violation).

As already mentioned in the introduction, the main point of breaking CP is
to get a matter-antimatter asymmetry. Indeed even if C is broken, this is
in general not enough to reach this goal. Indeed any other symmetries
involving C (like CP, but CS in general) leads to matter-antimatter
symmetry. In 6D the C symmetry is automatic for gauge interactions and the
symmetry particle/antiparticle is respected. In 4D C is not automatic but
CP leads to the same conclusion. Our idea to break this symmetry is
precisely to introduce a compactification which breaks all these CS
symmetries.

\section{Hosotani mechanism with two ED}\label{sec:Hosotani}

At the moment we work on flat space-time $\mathbb{M}^4\times
\mathbb{R}^2/G$ where the two ED are compactified by means of
orbifolding\footnote{In this note, \textit{"orbifold"} refers to any
quotient spaces regardless of the existence of fixed points.} through one
of the 17 two-dimensional space groups $G$ \cite{Nilse:2006}. These groups
correspond to isometries of $\mathbb{R}^2$, which include translations,
$2\pi/n$-rotations ($n=2,3,4$ and $6$), reflections and glide
reflections\footnote{Translations combined with mirror reflection.}. These
isometries must obviously be symmetries of the 6D original lagrangian. For
instance, only translations and rotations can be used with a chiral
lagrangian, and the possible orbifolds in this case are\footnote{The flat
torus $T^2$ has no fixed point and is generally not called orbifold.}:
$T^2$, $T^2/\mathbb{Z}_2$, $T^2/\mathbb{Z}_3$, $T^2/\mathbb{Z}_4$ and
$T^2/\mathbb{Z}_6$. We will see later that such lagrangians lead to highly
non trivial issues which are due to chiral anomalies and to the
interpretation of quantum corrections in ED models. Until then, we will
nevertheless stick to them.

In any case, two kinds of compactification exist: the "non-magnetized" and
the "magnetized" one. In the first case, a non zero field strength is
unstable and the only solutions are flat connections. In the second case,
a non zero field strength can be stable and the solution corresponds to a
physical flux orthogonal to the ED. The stability is ensured by the
quantization of the flux for topological reasons \cite{Alfaro:2007}.

Let us focus on the flat torus $T^2$ characterized by two radii $R_4$ and
$R_5$ (we don't consider here issues of gravitational stability, and they
are seen as free parameters). Because of the translation symmetry on the
torus, gauge fields on this manifold must be periodic up to a gauge
transformation \cite{Alfaro:2007}\cite{Salvatori:2007}: \beq A_a(y+2\pi
R_i)=T_i(y)A_a(y)T_i^{-1}(y)+\frac{i}{g}T_i(y)\partial_a T_i^{-1}(y)
\label{BCT2}
\eeq The topology of the torus requires\footnote{This is true if we
introduce fermions in a representation sensitive to the center of the
group (\textit{e.g.} the fundamental one). However, as long as we work
with insensitive representations, the relation is valid up to an element
of the center of the group\cite{Alfaro:2007}\cite{Salvatori:2007}. We
neglect this at the moment.} $T_4(y+2\pi R_5)T_5(y)=T_5(y+2\pi
R_4)T_4(y)$.

However we must be careful, because the BC, $T_i$, do not fix the symmetry
of the effective 4D theory. Indeed, the component of the gauge fields in
the ED, playing the role of scalar fields in 4D could very well acquire a
\textit{"vev"} through quantum effects. More precisely, the ED space being
multiply-connected, some non-integrable phase factors become dynamical
variables which can lead to effective symmetry breaking in 4D. Indeed,
it's worth stressing that neither \textit{"vev"} nor BC are gauge
invariant concepts. The true gauge invariant quantities are the so called
\textit{Wilson lines phases} defined by Hosotani as the eigenvalues of
$W_{C_i}(y)T_{C_i}$, with: \beq W_{C_i}(y)=\mathcal{P} \
\exp\left(ig\int_{C_i}\mathrm{d}y'_j \langle A_j(y')\rangle
\right),
\label{WLph}
\eeq for all the non equivalent non-contractible cycles $C_i$ starting at
$y$, and $T_{C_i}$ the associated BC.

In the following we will restrict ourselves to SU(N) gauge groups for
which we have an important result \cite{Alfaro:2007}\cite{Salvatori:2007}:
because of the non existence of topological quantities on $T^2$, all
stable configurations correspond to flat connexions $\langle F_{45}\rangle
=0$. In his approach \cite{Hosotani:1989}, Hosotani takes this result as
an hypothesis. Moreover, he restricts himself to homogeneous BC,
\textit{i.e.} $T_i(y)=T_i$. This is not mandatory, but it can help somehow
to get a better insight of the physics. For this reason we first give a
quick analysis of the simple case, followed by a more general, but also
more technical one.

To elucidate the 4D symmetry, we are particularly interested in the zero
modes ($y$ independent) of the gauge field. Obviously these correspond to
directions in the gauge group which remain unbroken after the
compactification. $F_{45}=0$ makes them satisfy $[\langle
A_4\rangle,\langle A_5\rangle]=0$. The homogeneous BC add the constraints
$[\langle A_i\rangle,T_j]=0$. In other words, $\langle A_4\rangle$ and
$\langle A_5\rangle$ must be part of the Cartan subalgebra of the group.
The selection of a particular solution is done at the quantum level
through the so called Hosotani mechanism. Therefore, we need to compute
the effective potential for $A_i$ to find the physical symmetry. The
result is of course affected by the geometry and the matter content (see
section \ref{sec:CPviolbyCB}).

The \textit{"vev's"} $\langle A_4\rangle$ and $\langle A_5\rangle$ can be
gauged away by the transformation\footnote{Since $[\langle
A_4\rangle,\langle A_5\rangle]=0$, $\Omega$ can be decomposed into
$\Omega_4(y) \Omega_5(y)=\Omega_5(y) \Omega_4(y)$ with
$\Omega_a(y)=\exp\left[-ig \langle A_a\rangle y_a\right]$.}: \beq
\Omega(y)=\exp\left[-i g \left(\langle A_4\rangle y_4 + \langle A_5\rangle
y_5\right)\right],
\label{omega}
\eeq and the BC matrices $T_i$ then become\footnote{$\Omega(-2\pi R_i)$ is
a shorthand notation for $\Omega(-2\pi R_4,0)$ or $\Omega(0,-2\pi R_5)$.}
$T^{\text{sym}}_i=\Omega(-2\pi R_i) T_i$. As previously mentioned, neither
$T_i$ nor $\langle A_i\rangle$ are physical, but only an appropriate
combination. Dynamics with different $T_i$ will give different $\langle
A_i \rangle$, but the "symmetric" BC, $T^{\text{sym}}_i$, obtained when
the \textit{"vev's"} are gauged away, are all equal\footnote{This is not
true for all topologies. Indeed, it may be that some BC cannot be linked
by any gauge transformation (\ref{omega}) for topologically satisfactory
$\langle A \rangle$. It follows that we could have more than one
equivalence class for BC (see for example \cite{Hosotani:2004}). Here, any
$T_i$ can be written as $\Omega(-2\pi R_i)$ thanks to the commutation
properties and we have only one equivalence class.}. Therefore, in all
generality, we can choose $T_i=1$ at the beginning and compute the
\textit{"vev"} $\langle A\rangle^{\text{phys}}$ which contains all the
physics.

Let us consider now the case where $T_i(y)$ can be $y$ dependent. The
result $\langle F_{45}\rangle=0$ is still valid \cite{Alfaro:2007} and
therefore the vacuum configuration for $\langle A \rangle$ must be pure
gauge (this time we don't make any \textit{a priori} assumption about $y$
dependence of it): \beq \langle A_a(y) \rangle =\frac{i}{g}U(y)\partial_a
U^{-1}(y), \nn
\eeq where $U$ must be compatible with the BC. If we use this expression
for $\langle A_a \rangle$ into equation \ref{BCT2}, it is easy to show
that $U$ must satisfy $U(y+2\pi R_i)\partial_a U^{-1}(y+2\pi
R_i)=T_i(y)U(y)\partial_a \left( T_i(y)U(y)\right)^{-1}$ what means: \beq
U(y+2\pi R_i)=T_i(y)U(y)V_i^{-1},
\label{Ufunction}
\eeq with $V_i$ a constant element of the gauge group such that
$[V_4,V_5]=0$ because of the topology. For some given BC, all the
classical vacua can be found by solving (\ref{Ufunction}) for all possible
$V_i$. Since $\langle F_{45}\rangle=0$, we know that solutions must exist,
at least for some compatible $V_i$. Moreover, it can be shown
\cite{Alfaro:2007}\cite{Salvatori:2007} that, for SU(N) groups on $T^2$,
solutions exist for any compatible $V_i$. A particular vacuum is labelled
by $T_i(y)$ and $U(y)$. Now let us perform a gauge transformation
$U^{-1}$. Then $\langle A(y) \rangle = 0$ and $T_i(y)=V_i$. Therefore, all
possible classical vacua can be labelled by constant and commuting BC:
\beq V_i=\exp\left(i \Theta_i\right),\nn
\eeq where $\Theta_i$ are constant and commuting matrices of SU(N)
algebra. Again quantum effects select the true vacuum which depends on
geometry and matter content (see section \ref{sec:CPviolbyCB}). Let us
call it $\Theta_i^{\text{phys}}$. After the gauge transformation: \beq
\Omega'(y)=\exp\left[-i\left(\frac{\Theta_4^{\text{phys}}}{2\pi
R_4}y_4+\frac{\Theta_5^{\text{phys}}}{2\pi R_5}y_5\right)\right], \nn
\eeq we end up with trivial BC and a \textit{"vev"}
for the background that contains all the physics (as in the Hosotani
approach): \beq \langle A_i
\rangle^{\text{phys}}=\frac{\Theta_i^{\text{phys}}}{2\pi g R_i}.\nn
\eeq This last identification is correlated by the computation of the WL
phases (\ref{WLph}) in the two approaches. In the first one with trivial
BC we find $W_i=\exp(i 2\pi g R_i \langle A_i^{\text{phys}} \rangle)$,
while in the second with trivial \textit{"vev"} we find $W_i=\exp(i
\Theta_i^{\text{phys}})$. Note also that it shows that the natural scale
for the effective \textit{"vev"} are the dimensions of the ED. This is
expected since they are the only dimensionfull parameters.

\section{CP violation induced by BC}\label{sec:CPviolbyCB}

In the last section we saw that the BC alone are not meaningful by
themselves. On the other hand, we can always perform a gauge
transformation that puts all the physics in the BC (in this gauge BC are
identified with the WL). In all that follows we will work in this gauge.
Therefore the fermionic fields\footnote{The notation refers explicitly to
the fundamental representation, but it can be easily extended to the
adjoint or others.} have the BC (to simplify notation $\Theta_i$ is
identified with $\Theta_i^{\text{phys}}$ defined in section
\ref{sec:CPviolbyCB}): \beq \Psi(y+2\pi R_i)=\exp(i \beta_i)
\exp\left(i\Theta_i\right) \Psi(y), \nn
\eeq with additional phases $\beta_i$ allowed because fermions appear
always in bilinears\footnote{We stress again that matter content plays a
crucial role in the dynamics that selects the physical vacuum at quantum
level. At this point, we suppose this vacuum known and encoded in the
BC.}. Note that $\beta_i$ phases (or Scherk-Schwarz (SchSch) phases) are
free external parameters and that we can choose them different for each
fermionic field. They will enter the dynamics of fermion, possibly
creating masses.

To study whether or not these BC lead to CP violation at the 4D level, we
need to check their compatibility with the transformations $\Psi
\rightarrow \Psi^*$ and $Y \rightarrow \hat{Y}$ (see section
\ref{sec:PCCP}). Remember that $\hat{Y}$ can be any rotation of
$(y_4,-y_5)$ (see the transformation (\ref{CP4d6d}) which makes explicit
the link between C in 6D and CP in 4D) and that CP is conserved in 4D as
long as we can find compatibility for one rotation. Here the gauge
symmetry adds an additional freedom. Indeed, the transformations can be
$\Psi \rightarrow U^* \Psi^*$, where $U$ is any global symmetry matrix
(since it keeps $\langle A\rangle=0$).

Under the prescribed symmetries, the two BC become\footnote{It may be
surprising that $\langle A \rangle$ doesn't change under the rotations.
One can understand that if one remembers that the physical quantities are
WL which are of course rotationally invariant.}: \beqn
\Psi^{\text{CP}}(y_4+2\pi R_4 \cos \theta, y_5 +2\pi R_4 \sin \theta) &=&
\exp\left[-i\beta_4\right]\exp\left[-i  \left(U \Theta_4 U^{-1}\right)^*
\right] \Psi^{\text{CP}}(y_4,y_5) \nn \\
\Psi^{\text{CP}}(y_4+2\pi R_5 \sin \theta, y_5 -2\pi R_5 \cos \theta)
&=& \exp\left[-i\beta_5\right]\exp\left[-i \left(U\Theta_5 U^{-1}\right)^*
\right] \Psi^{\text{CP}}(y_4,y_5) \nn
\eeqn The Table \ref{tab:SymCP}
shows the different symmetries which might be compatible with BC. The
angle $\theta$ refers to the rotation $\mathcal{R}$. The columns marked
$\beta_4$ and $\beta_5$ indicate a possible constraint for these phases.
The next two columns show the constraints on the $U$ matrix introduced
above\footnote{We should write $\left(U \Theta U^{-1}\right)\sim
\pm\Theta^*$, but remember that $\Theta_a^{\dagger}=\Theta_a$, then we can
use $\Theta_a^T$ instead of $\Theta_a^*$. However, $\Theta_a$'s are
diagonal (or can be diagonalized because of the topology), and therefore
we can use $\Theta_a$. Note also that these relations are not so strict.
Indeed the periodicity of the exponential factor must be taken into
account.}. Note that for adjoint fermions, insensitive to the centre of
the group, we have a little bit more freedom. The $k$ and $k'$ factors
take this into account for SU(N) groups ($T=\text{diag}(1,...,1,1-N)$).
$k$ and $k'$ can take all integer values for representations which are
insensitive to the centre, but must be zero in the other case.
\begin{table}[h!]
\centering
  \begin{tabular}{|c|c||c|c|c|c|c||c|}
\cline{3-7} \multicolumn{2}{c||}{} & $\theta$ & $\beta_4$ & $\beta_5$ & $U
\Theta_4 U^{-1}$ & $U \Theta_5 U^{-1}$ & \multicolumn{1}{c}{} \\
  \hline\hline &  & $0$      & $\left\lbrace 0,\pi\right\rbrace$ &
  $\left[0,2\pi\right[$ & $-\Theta_4+\frac{2\pi k}{N}T$          &
  $\Theta_5+\frac{2\pi k'}{N}T$ & (1)
 \\
\cline{3-8} &  $R_4\neq R_5$ & $\pi$    & $\left[0,2\pi\right[$ &
  $\left\lbrace 0,\pi\right\rbrace$ & $\Theta_4+\frac{2\pi k}{N}T$ &
  $-\Theta_5+\frac{2\pi k'}{N}T$ & (2) \\
\cline{2-8} \multicolumn{2}{|c||}{}   & $\pi/2$  & $-\beta_5$ &
  $-\beta_4$ & $-\Theta_5+\frac{2\pi k}{N}T$          &
  $-\Theta_4+\frac{2\pi k'}{N}T$ & (3) \\
\cline{3-8} \multicolumn{2}{|c||}{$R_4=R_5$}               & $3\pi/2$ &
  $\beta_5$ & $\beta_4$ & $\Theta_5+\frac{2\pi k}{N}T$ &
  $\Theta_4+\frac{2\pi k'}{N}T$ & (4) \\
  \hline
  \end{tabular}
  \caption{Hypothetical transformations that could be identified with an
  effective CP symmetry in 4D if compatible with boundary conditions (BC).
  \label{tab:SymCP}
  }
  \end{table}

We may expect a large variety of situations depending of the gauge group.
  Let us look here to some simple examples in SU(2), which we are
  particularly interested in (see section \ref{sec:Examples}). We always
  have $\Theta_4=a t_3$ and $\Theta_5=b t_3$ ($t_3=\sigma_3/2$).
  Therefore, if the constraints on $\beta$'s and radii are fulfilled: the
  transformations $(1)$ and $(2)$ are good candidates for CP symmetry
  either if $a$ or $b=(j+k/2)\pi$, while the transformations $(3)$ and
  $(4)$ are good candidates either if $a+b$ or $a-b=(j+k/2)\pi$. $k$ and
  $j$ are integer numbers. $j$ can always be non zero because it stands
  for the periodicity in the exponential factor, but $k$ can only be non
  zero for representations insensitive to the centre.

There are now two main questions. (1) Which patterns can be realized (and
under which conditions)? (2) At which level does CP violation manifest
itself (and what could be phenomenologically promising)? As mentioned in
the introduction, answering the first one is tricky because we need to
compute the effective potential for WL for each group we want to study and
then find the minima of this potential which depend on many parameters
(SchSch phases, radii ratio, matter content). While the case of SU(2) on
$S^1$ has been extensively studied, the behaviour for larger groups on
$T^2$ becomes quickly hard to discuss. For the time being we focus
ourselves here on simple examples. Regarding the problem of phenomenology,
one of the main limitations (without any new mechanism) has been mentioned
and concerns the absence of gap between light and heavy sectors. A partial
answer to this issue (unfortunately quite inelegant) comes from the SchSch
phases. If we choose them sufficiently small, they could account for small
masses of the previously massless modes. We must however remember their
influence on the dynamics of WL.

CP violation is, even in the Standard Model, a tricky issue to characterize
(the Jarlskog determinants providing a partial answer). To prove that CP
is violated, the safest way is to provide an "observable". Here we will
deal with a single (light) fermion species and the simplest "observable"
is then the electric dipole moment (EDM) of the lightest mode\footnote{We
study the lightest mode since we look for an understanding of CP violation
at low energy. However a zero EDM for this state doesn't mean that CP is
conserved (and that our previous analysis fails), as it may manifest
itself at higher energy. Remember also that an EDM violates both P and CP.
It is however easy to check that, with this mechanism, the 4D P symmetry
is broken as soon as the CP one is.}.\\

\section{Examples with SU(2)}\label{sec:Examples}

For the next examples, we will work with one of the simplest groups,
\textit{i.e.} SU(2). In the two first examples the matter content consists
in a fermion in the fundamental (resp. the adjoint) representation. In
SU(2) there are two independent dynamical variables called $\theta_4$ and
$\theta_5$ such that $\Theta_a =
\begin{pmatrix}
\theta_a & 0 \\
0 &-\theta_a
\end{pmatrix}
.$

The effective potential can be decomposed into \cite{Hetrick:1989}: \beq
V_{\textbf{eff}}=V\left(-V_{\textbf{eff}}^{\text{g+gh}}+\sum_i
2V_{\textbf{eff}}^{\text{f}_i}+\sum_i 2V_{\textbf{eff}}^{\text{ad}_i}
\right),\nn
\eeq where $V$ is a positive constant, $V_{\textbf{eff}
}^{\text{g+gh}}$ the contribution from gauge and ghost fields,
$V_{\textbf{eff}}^{\text{f}_i}$ the contribution from fundamental fermions
and $V_{\textbf{eff}}^{\text{ad}_i}$ the one from adjoint fermions. Each
contribution can be written as an infinite sum over fields modes. It worth
noting that this expression is only valid for Dirac spinors, and not Weyl
spinors. From now on, we will use it nonetheless, and postpone the
justification to the next section.

The potential must be studied numerically. The results for a theory with
only fundamental fermions are simple and given in Table \ref{tab:SU2Fund}.

\begin{table}[h!]
\centering
\begin{tabular}{|l||c|c|c|c|}
\cline{2-5} \multicolumn{1}{l||}{} & $\beta_5 \in [0,\pi/2]$ & $\beta_5
\in [\pi/2,\pi]$ & $\beta_5 \in [\pi,3\pi/2]$ & $\beta_5 \in
[3\pi/2,2\pi]$ \\
\hline\hline $\beta_4 \in [0,\pi/2]$ & $(\pi,\pi)$ & $(\pi, 0 )$ &
$(\pi, 0 )$ & $(\pi,\pi)$ \\
\hline $\beta_4 \in [\pi/2,\pi]$ & $( 0 ,\pi)$ & $( 0 , 0 )$ & $( 0 , 0
)$ & $( 0 ,\pi)$ \\
\hline $\beta_4 \in [\pi,3\pi/2]$ & $( 0 ,\pi)$ & $( 0 , 0 )$ & $( 0 ,
0 )$ & $( 0 ,\pi)$ \\
\hline $\beta_4 \in [3\pi/2,2\pi]$ & $(\pi,\pi)$ & $(\pi, 0 )$ & $(\pi,
0 )$ & $(\pi,\pi)$ \\
\hline
\end{tabular}
\caption{Wilson line (WL) phases for a SU(2) theory with a 6D spinor in
the fundamental representation.
\label{tab:SU2Fund}
}
\end{table}

According to \cite{Hetrick:1989}, this result is valid for $R_4=R_5$, but
our study shows that this remains exact even for $R_4\neq R_5$. To be more
precise, the potential shape depends only on $r=R_5/R_4$.
When\footnote{The case $r<1$ is completely symmetric.} $r>1$, the
potential flattens in the $y_5$ direction, but the global minimum stays
unchanged at least for $r\lesssim 5$. Beyond, an other local minimum
becomes very close to the global one and it is hard to select the right
one with numerical calculations. Nevertheless, the two candidates lead to
the same phenomenological issues that we will describe here. First, it's
worth noting that $\theta=0$ and $\theta=\pi$ are particular values since
then $-\theta=\theta$, $\exp[i\Theta_a]=\pm\bs{1}$ and the gauge symmetry
remains unbroken because all SU(2) generators commute with transition
functions $T_i$. However CP symmetry can still be broken because of the
SchSch phases or $R_4\neq R_5$ (see Table \ref{tab:SymCP}), but another
big issue is the absence of a light fermion, even with $\beta$'s tuned to
be small. Indeed, when $\beta$'s are small, the WL are large and vice
versa. More precisely one can show that the smallest "distance" between
$(\beta_i+\theta_i)/2\pi$ and an integer is $0.25$. Then the fermion
masses are bounded from below (with $R_4=R_5=R$) $m_f > \sqrt{2}/4R\sim
0.35/R$ and there is a poor gap between the lightest mode and the KK
tower.

%

Let us focus now on a more interesting example. Richer phenomenology can
be reached if we replace the fundamental fermion by an adjoint. We will
not try to give an exhaustive study of the effective potential in this
case. Refs and personal analysis show that, at least in the interesting
regime $\beta_4,\beta_5 \in [0,0.1]$ and $0.9<r=R_5/R_4<1$,
$(\theta_4,\theta_5)=(\pi/2,\pi/2)$. This is interesting because this time
the SU(2) symmetry is spontaneously broken into U(1), and after this
breaking we have a neutral fermion with mass $\sim \beta/\sqrt{2}\pi R$
which can be choose to be small. Moreover Table \ref{tab:SymCP} tells us that
CP can be broken with non zero $\beta$'s. If $r=1$, $\beta_4$ must be
different from $\beta_5$, but if $r\neq 1$ this is not even necessary. We
will verify these affirmations with the EDM of our light fermion. Details
about particle content and effective interactions can be found in the
Appendix \ref{AppA}. The EDM is given by\footnote{We normalize to the
scale $R$ and the coupling constant $e$ of the SU(2) gauge interaction.}:
\beqn \left|\frac{d_E R}{e^3}\right| &=& \left| \sum_{nm}\left\lbrace
F^+_{nm}\sin(\varphi_{3;00}-\varphi_{+;nm})+J^+_{nm}\sin\varphi_{3;00}
+K^+_{nm}\cos\varphi_{3;00}\right\rbrace\right. \nn \\
&+& \left. \sum_{nm}\left\lbrace
F^-_{nm}\sin(\varphi_{3;00}-\varphi_{-;-n-m})+J^-_{nm}\sin\varphi_{3;00} +
K^-_{nm}\cos\varphi_{3;00} \right\rbrace \right|,
\label{EDMana}
\eeqn where the coefficients $F$, $J$ and $K$ and the phases
$\varphi_{\pm;nm}$ and $\varphi_{3;nm}$ are functions of the $\theta$'s,
the $\beta$'s and $r$. Their explicit form can be found in the Appendix
\ref{AppB}. When $r=1$ and $\beta_4=\beta_5$, it is easy to check that
(see Appendix \ref{AppB}): \beq F^{\pm}_{nm}= F^{\pm}_{mn} ; \quad
J^{\pm}_{nm}= -K^{\pm}_{mn} ; \quad
\varphi_{\pm;nm}=\frac{\pi}{2}-\varphi_{\pm;mn}; \quad
\varphi_{3;00}=\frac{\pi}{4}(3-2\ \text{sign}(\beta)) .
\label{relBetaEq}
\eeq Therefore: \beq \left|\frac{d_E R}{e^3}
\right| \sim(\sin\varphi_{3;00}-\cos\varphi_{3;00})=0.\nn
\eeq This is no more true when $r\neq 1$ or $\beta_4\neq \beta_5$. We will
illustrate this with numerical evaluations. Our results can be found in
Table \ref{tab:SU2edm}. We use the notation $\beta=\beta_4$,
$\Delta\beta=\beta_4-\beta_5$, $\Delta r=1-r$.

\begin{table}[h!]\centering
\begin{tabular}{|c|c|c|c|c|}
\hline
$\beta$   & $\Delta\beta/\beta$ & $\Delta r$ & $m_{\text{light}}R$ & $d_E R/e^3$  \\ \hline \hline
$[0,10^{-1}]$ & 0 & 0 & $\sqrt{2}\beta$ & $0$ \\ \hline
$10^{-1}$ & $0$           & $10^{-1}$  & $1.35 \ 10^{-1}$    & $1.09 \ 10^{-3}$ \\ \hline
$10^{-1}$ & $0$           & $10^{-2}$  & $1.41 \ 10^{-1}$    & $0.99 \ 10^{-4}$ \\ \hline
$10^{-1}$ & $0$ 		  & $10^{-3}$  & $1.41 \ 10^{-1}$    & $0.98 \ 10^{-5}$ \\ \hline
$10^{-1}$ & $0$ 		  & $10^{-4}$  & $1.41 \ 10^{-1}$    & $0.98 \ 10^{-6}$ \\ \hline
$10^{-1}$ & $10^{-1}$     & $0$        & $1.35 \ 10^{-1}$    & $4.66 \ 10^{-3}$ \\ \hline
$10^{-1}$ & $10^{-2}$     & $0$        & $1.41 \ 10^{-1}$    & $4.50 \ 10^{-4}$ \\ \hline
$10^{-1}$ & $10^{-3}$     & $0$        & $1.41 \ 10^{-1}$    & $4.48 \ 10^{-5}$ \\ \hline
$10^{-1}$ & $10^{-4}$ 	  & $0$        & $1.41 \ 10^{-1}$    & $4.48 \ 10^{-6}$ \\ \hline
$10^{-2}$ & $10^{-1}$ 	  & $0$        & $1.35 \ 10^{-2}$    & $4.28 \ 10^{-3}$ \\ \hline
$10^{-3}$ & $10^{-1}$ 	  & $0$        & $1.35 \ 10^{-3}$    & $4.28 \ 10^{-3}$ \\ \hline
$10^{-3}$ & $10^{-1}$ 	  & $10^{-1}$  & $1.27 \ 10^{-3}$    & $5.71 \ 10^{-3}$ \\ \hline
$10^{-3}$ & $10^{-1}$ 	  & $10^{-2}$  & $1.33 \ 10^{-3}$    & $4.41 \ 10^{-3}$ \\ \hline
$10^{-3}$ & $10^{-1}$ 	  & $10^{-3}$  & $1.34 \ 10^{-3}$    & $4.29 \ 10^{-3}$ \\ \hline \hline
\end{tabular}
\caption{Numerical evaluation of the electric dipole moment (EDM) of
$\psi_{3;00}$ particle (in a SU(2) theory with a 6D spinor in the adjoint
representation) for different sets of parameters. The sum on $(n,m)$ are
limited to $n,m \in [-10;10]$, and we find no big deviation from results
obtained with $n,m \in [-50;50]$. We give also its mass $m_{3;00}$ which
is the lightest of the fermion spectrum. See text for more precision.
\label{tab:SU2edm}
}
\end{table}

The behaviour of the lightest mass is easily predicted. Indeed (see
Appendix \ref{AppA}) we have $m_{\text{light}}R\simeq \beta
(1+\Delta\beta/\beta+\Delta r)$, and its order of magnitude is directly
related to $\beta$. On the other hand, the behaviour of the EDM is less
intuitive from the analytic solutions, because of the summations and
integrations in its expression. Nevertheless, we could expect a behaviour
of the type: \beq \left|\frac{d_E R}{e^3}\right| \simeq C \cdot
\left(\Delta r + \kappa \frac{\Delta \beta}{\beta}\right),\nn
\eeq where $C$ and $\kappa$ are (almost) constant factors. Numerical
evaluations show this is the case (with a pretty good accuracy) with
$C\sim 10^{-2}$ and $\kappa \sim 4.5$. Obviously, the dominant CP source
($\Delta r$ or $\Delta \beta/\beta$) dictates the order of magnitude for
the EDM.

\section{Chiral anomaly in 6D} \label{sec:ChAno6D}

Gauge theories in more than 4 space-time dimensions are not renormalizable
and it could then seems dangerous to consider quantum corrections (in the
effective potential for instance) in this context. However, the
WL-dependent part of $V_{\textbf{eff}}$ turns out finite, at least at the
one loop level, and can then be evaluated unambiguously
\cite{Hosotani:2005}. For chiral theory, however, things become worse,
because of the presence of anomalies. In 6D, they come from the square
diagram\footnote{In non abelian theories, there exists other pathological
diagrams, but they can be related to this one through gauge invariance.}
(equivalent to triangle diagram in 4D) which certainly plays a role in the
effective potential.

Concretely, anomalies originate from UV divergences, but they are finite
and calculable IR effects (even in more than 4D) which then do not depend
on the UV completion of the theory
\cite{Frampton:1983}\cite{Scrucca:2004}. For non renormalizable theories,
which are only valid under a certain energy scale, anomalies can cancel
among themselves (like in 4D), but they can also cancel with effects
originating from an unknown UV sector. It is not our point to discuss
these issues here, and we will avoid them with the introduction of both 6D
chiralities ($+$ and $-$) in the same representation. However, we will use the BC to differentiate the masses of the light excitations of these fields. As announced, this
will lead to small modifications in the effective potential, but our
previous results will remain intact.

The only modification of $V_{\textbf{eff}}$ appears in the fermion
contributions. We must do the replacements :
\beqn 2 V_{\textbf{eff}}^{\text{f}_i}(\beta_4,\beta_5) &\longrightarrow &
V_{\textbf{eff}}^{\text{f}_i}(\beta_{4+},\beta_{5+})+
V_{\textbf{eff}}^{\text{f}_i}(\beta_{4-},\beta_{5-}) \nn \\
2 V_{\textbf{eff}}^{\text{ad}_i}(\beta_4,\beta_5) &\longrightarrow &
V_{\textbf{eff}}^{\text{ad}_i}(\beta_{4+},\beta_{5+})+
V_{\textbf{eff}}^{\text{f}_i}(\beta_{4-},\beta_{5-}).\nn
\eeqn We have checked numerically that, in the range of SchSch phases we
work with, this keeps the minimum of the total effective potential at
$(\theta_4,\theta_5)=(\pi/2,\pi/2)$. This is why we used it in section
\ref{sec:Examples}, even if at that time, we had only introduced one
chirality\footnote{Note by the way that the same conclusion holds for
fundamental representation and our previous results stay qualitatively
identical. There is no way to force $\theta$ values to be different from
$0$ or $\pi$ which lead to unbroken gauge symmetry. Moreover, the minimum
always arranges to prevent small masses in the spectrum.}.

It is easy to verify that in the case of degenerate SchSch phases for $+$
and $-$ chiralities, the EDMs are exactly opposite for the two sectors,
thus restoring CP ("CP doubling"). To be more precise, the lightest modes
could still be distinguished through their different couplings, but we
prefer to provide a case where CP violation is explicit in terms of low
energy parametrization. This is easily obtained if we choose different
SchSch phases for $+$ and $-$ chiralities. As a simple class of examples,
let us take $\Delta r=0$, $\Delta \beta_-=0$ and $\Delta \beta_+\neq 0$.
In this way, in the "$-$" sector $d_E=0$ while in the "$+$" sector
$d_E\neq 0$.

\section{Conclusion and perspectives}\label{sec:Conclusion}

We made use of the Hosotani mechanism to generate both gauge and CP
symmetry breaking through compactification from a 6-dimensional model.
Though we found examples where it works, our solutions is far from being
realistic, and they must be seen more as \textit{"proof of concept"}. One
of the major difficulty of the work is the high level of entanglement in
the approach. Indeed, the final result depends both on matter content
(representations), BC (SchSch phases) and WL phases, while the latter
depend in turn on the formers and are dynamically determined through a
potential which must be numerically evaluated.

The next steps in this program should be the resolution of the two main
drawbacks of the present solutions. First new compactification mechanism
(like orbifold or flux compactification\footnote{See for instance interesting application for gauge symmetry breaking in \cite{Faedo:2010}.}) might be employed to reach a
chiral theory in 4D (at this point the only difference between left and
right couplings in the gauge sector comes through a phase). Moreover, we'd
like to avoid the presence of two (nearly) identical fermionic sectors
without introducing anomalies in the theory. Secondly (but this maybe even
more ambitious), a mechanism which produces a low energy sector naturally
separated from the Kaluza-Klein scale would be very welcome. For instance,
in more complex situations, one can hope for an effective low energy
potential between the remaining scalars, what would provide the lower mass
scale, but this goes beyond this "proof of concept" paper.

\section*{Acknowledgments} 

We are indebted to  E.~Nugaev and S.~Troitsky for helpful discussions. M.L. thanks the Service de Physique Th\'{e}orique at Universit\'{e} Libre de Bruxelles for kind hospitality. This work is funded in part by IISN and by Belgian Science Policy (IAP VII/37). This work has been supported in part by the grant of the President of the Russian Federation NS-2835.2014.2 (M.L.).

\appendix

\section{Dirac matrices in 6D}\label{AppA1}

We use the following representation for the Dirac matrices in 6D: \beq
\Gamma^A=
\begin{pmatrix}
0 & \Sigma^A \\
\bar{\Sigma}^A & 0
\end{pmatrix}
\ \ \text{and} \ \ \ \Gamma^7=
\begin{pmatrix}
1 & 0 \\
0 & -1
\end{pmatrix}
, \nn
\eeq where $\Sigma^{\mu}
=\gamma^0\gamma^{\mu}$, $\Sigma^4=i\gamma^0\gamma^5$,
$\Sigma^5=\gamma^0$ and $\bar{\Sigma }^{0}=\Sigma ^{0}$, $\bar{\Sigma
}^{A\neq 0}=-\Sigma ^{A\neq 0}$. The $\gamma$'s are 4D Dirac matrices: \beq
\gamma^{\mu} =
\begin{pmatrix}
0 & \sigma_{\mu} \\
\bar{\sigma}_{\mu} & 0
\end{pmatrix}
\ \ \text{and} \ \ \ \gamma^5 =
\begin{pmatrix}
1 & 0 \\
0 & -1
\end{pmatrix}
,\nn
\eeq where $\sigma_{\mu}
=(1,\sigma_i)$ and $\bar{\sigma}_{\mu}=(1,-\sigma_i)$.

\section{Effective 4D theory for an SU(2) adjoint fermion}\label{AppA}

Not considering here the anomalies, we work only with a 6D Weyl fermion
$\Psi$ in the adjoint representation of SU(2) and a gauge field $A_A$.
These fields can be decomposed in the Cartan basis $\left\lbrace
T_+,T_-,T_3\right\rbrace$ which satisfies $[T_+,T_-]=T_3$ and
$[T_2,T_{\pm}]=\pm T_{\pm}$. The 6D lagrangian can be written \beq
\mathcal{L}=-\frac{1}{2}\text{Tr}[F_{AB}F^{AB}]+2 \ \text{Tr}[i
\Psi^{\dagger}\bar{\Sigma}^A D_A \Psi],\nn
\eeq with $F_{AB}
=\partial_A A_B-\partial_B A_A-ie[A_A,A_B]$, $D_A=\partial_A-i e
[A_A,\bullet]$ the covariant derivative and
$\bar{\Sigma}^A=\gamma^0\cdot\left\lbrace \gamma^0, -\gamma^i, -i\gamma^5,
-1\right\rbrace$. If we define $\psi=\gamma^0 \Psi$ we can write the
fermionic part of the lagrangian in the following form:
\begin{align}
\mathcal{L}&\supset\ i \bar{\psi}_+\gamma^{\mu}\partial_{\mu}\psi_+ +i
\bar{\psi}_-\gamma^{\mu}\partial_{\mu}\psi_- +i
\bar{\psi}_3\gamma^{\mu}\partial_{\mu}\psi_3 \nn \\
&-i \bar{\psi}_+(\partial_5-i\gamma^5\partial_4)\psi_+ -i
\bar{\psi}_-(\partial_5-i\gamma^5\partial_4)\psi_- -i
\bar{\psi}_3(\partial_5-i\gamma^5\partial_4)\psi_3 \nn \\
&+e\bar{\psi}_+ \gamma^{\mu}(A_{3;\mu}\psi_+ - A_{+;\mu}\psi_3)
+e\bar{\psi}_- \gamma^{\mu}(A_{-;\mu}\psi_3 - A_{3;\mu}\psi_-)
+e\bar{\psi}_3 \gamma^{\mu}(A_{+;\mu}\psi_- - A_{-;\mu}\psi_+)\nn \\
&-e\bar{\psi}_+ (A_{3;5}\psi_+ - A_{+;5}\psi_3) -e\bar{\psi}_-
(A_{-;5}\psi_3 - A_{3;5}\psi_-) -e\bar{\psi}_3 (A_{+;5}\psi_- -
A_{-;5}\psi_+)\nn\\
&+e\bar{\psi}_+ i\gamma^{5}(A_{3;4}\psi_+ - A_{+;4}\psi_3)
+e\bar{\psi}_- i\gamma^{5}(A_{-;4}\psi_3 - A_{3;4}\psi_-) +e\bar{\psi}_3
i\gamma^{5}(A_{+;4}\psi_- - A_{-;4}\psi_+)
\label{LfermiEff}
\end{align}
To find the 4D effective lagrangian we need to decompose $\psi$ and $A_A$
into fundamental modes which satisfy BC. For an adjoint fermion these are
given by: \beq \psi(y+2\pi R_i)=e^{i\beta_i}e^{i\theta_i
T_3}\psi(y)e^{-i\theta_i T_3},\nn
\eeq or, in the Cartan basis: \beqn \left\{ \begin{aligned}
&\psi_3(y+2\pi R_i)=e^{i\beta_i}\psi_3(y)\\
&\psi_{\pm}(y+2\pi R_i)=e^{i(\beta_i\pm \theta_i)}\psi_{\pm}(y).\\
\end{aligned}
\right.\nn
\eeqn Therefore the (normalized) mode decompositions are: \beqn \left\{
\begin{aligned}
&\psi_3(y)=\frac{1}{2\pi\sqrt{R_4R_5}}\sum_{nm}e^{i\left(n+\frac{\beta_4}{2\pi}\right)\frac{y_4}{R_4}}e^{i\left(m+\frac{\beta_5}{2\pi}\right)\frac{y_5}{R_5}}\psi_{3;nm} \\
&\psi_{\pm}(y)=\frac{1}{2\pi\sqrt{R_4R_5}}\sum_{nm}e^{i\left(n+\frac{\beta_4\pm\theta_4}{2\pi}\right)\frac{y_4}{R_4}}e^{i\left(m+\frac{\beta_5\pm\theta_5}{2\pi}\right)\frac{y_5}{R_5}}\psi_{\pm;nm}.\\
\end{aligned}
\right.\nn
\eeqn The decompositions for $A_A$ are obtained with $\beta_4=\beta_5=0$.

Let us introduce these decompositions in (\ref{LfermiEff}). The first line
gives the kinetic energy for each mode. The second line gives the
effective 4D masses: \beqn \left\{
\begin{aligned}
&m_{3;nm}=-\frac{1}{R}\left[\left(m+\frac{\beta_5}{2\pi}\right)-i\gamma^5
r \left(n+\frac{\beta_4}{2\pi}\right)\right]\\
&m_{\pm;nm}=-\frac{1}{R}\left[\left(m+\frac{\beta_5\pm\theta_5}{2\pi}\right)-i\gamma^5 r \left(n+\frac{\beta_4\pm\theta_4}{2\pi}\right)\right].\\
\end{aligned}
\right.\nn
\eeqn To get real (and positive) masses, we perform a chiral rotation
$\psi\rightarrow e^{i\frac{\varphi}{2}\gamma^5}\psi$, where the phases are
given by: \beqn \left\{
\begin{aligned}
&\exp[i\varphi_{3;nm}]=-\frac{\left(m+\frac{\beta_5}{2\pi}\right)+ir\left(n+\frac{\beta_4}{2\pi}\right)}{\sqrt{\left(m+\frac{\beta_5}{2\pi}\right)^2+r^2\left(n+\frac{\beta_4}{2\pi}\right)^2}}\\
&\exp[i\varphi_{\pm;nm}]=-\frac{\left(m+\frac{\beta_5\pm\theta_5}{2\pi}\right)+ir\left(n+\frac{\beta_4\pm\theta_4}{2\pi}\right)}{\sqrt{\left(m+\frac{\beta_5\pm\theta_5}{2\pi}\right)^2+r^2\left(n+\frac{\beta_4\pm\theta_4}{2\pi}\right)^2}}.\\
\end{aligned}
\right.
\label{Cphases}
\eeqn The real masses are then: \beqn \left\{ \begin{aligned}
&m_{3;nm}=\frac{1}{R}\sqrt{\left(m+\frac{\beta_5}{2\pi}\right)^2+r^2\left(n+\frac{\beta_4}{2\pi}\right)^2}\\
&m_{\pm;nm}=\frac{1}{R}\sqrt{\left(m+\frac{\beta_5\pm\theta_5}{2\pi}\right)^2+r^2\left(n+\frac{\beta_4\pm\theta_4}{2\pi}\right)^2}.\\
\end{aligned}
\right.\nn
\eeqn

Note that the two last relations (\ref{relBetaEq}), valid for
$\beta_4=\beta_5$ and $r=1$ can be easily proven here with the definitions
(\ref{Cphases}). If we remind that the effective potential imposes
$\theta_4=\theta_5$, we see that the exchange of $n$ and $m$ in these
relations is equivalent to the exchange of real and imaginary part of the
phases, what means $\varphi_{\pm;nm}=\pi/2-\varphi_{\pm;mn}$. Finally
$\varphi_{3;00}=\pm \pi/2$, since in this case it has its real and
imaginary parts equal. The sign is determined by the sign of $\beta$, and
the solution can be written synthetically as
$\varphi_{3;00}=\frac{\pi}{4}(3-2\text{sign}(\beta))$.

The third line in (\ref{LfermiEff}) gives the effective interactions with
4D vector bosons, while the fourth and fifth ones give the interactions
with 4D scalars bosons. To get an interesting form we need to perform the
chiral rotation, but also to go in the mass eigenbasis for the bosons. To
study this let us have a look to the quadratic part of the gauge
lagrangian: \beqn \mathcal{L}&\supset &
-\frac{1}{4}\left(\partial_{\mu}A_{3;\nu}-\partial_{\nu}A_{3;\mu}\right)^2
+\frac{1}{2}(\partial_{\mu}A_{3;4})^2
+\frac{1}{2}(\partial_{\mu}A_{3;5})^2
+\frac{1}{2}(\partial_{4}A_{3;\mu})^2
+\frac{1}{2}(\partial_{5}A_{3;\mu})^2\nn \\
&-&\frac{1}{2}(\partial_{4}A_{3;5})^2
-\frac{1}{2}(\partial_{5}A_{3;4})^2
+(\partial_{4}A_{3;5})(\partial_{5}A_{3;4})
-(\partial_{4}A_{3;\mu})(\partial^{\mu}A_{3;4})
-(\partial_{5}A_{3;\mu})(\partial^{\mu}A_{3;5})\nn \\
&-&\frac{1}{2}\left|\partial_{\mu}A_{+;\nu}-\partial_{\nu}A_{+;\mu}\right|^2
+|\partial_{\mu}A_{+;4}|^2 +|\partial_{\mu}A_{+;5}|^2
+|\partial_{4}A_{+;\mu}|^2 +|\partial_{5}A_{+;\mu}|^2\nn \\
&-&|\partial_{4}A_{+;5}|^2 -|\partial_{5}A_{+;4}|^2
+\left[(\partial_{4}A^*_{+;5})(\partial_{5}A_{+;4})
-(\partial_{4}A^*_{+;\mu})(\partial^{\mu}A_{+;4})
-(\partial_{5}A^*_{+;\mu})(\partial^{\mu}A_{+;5})+h.c.\right]
\label{LgaugeEff}
\eeqn From BC we can convert $\partial_4$, $\partial_5$ into mass matrices
for the vector and scalar bosons. The vector bosons $A_{3;\mu,nm}$ (resp.
$A_{+;\mu,nm}$) have masses $M_{3;nm}$ (resp. $M_{+;nm}$) given by: \beqn
\left\{
\begin{aligned}
&M_{3;nm}=\frac{1}{R}\sqrt{m^2+r^2 n^2}\\
&M_{+;nm}=\frac{1}{R}\sqrt{\left(m+\frac{\theta_5}{2\pi}\right)^2+r^2\left(n+\frac{\theta_4}{2\pi}\right)^2}.\\
\end{aligned}
\right.\nn
\eeqn One of the bosons in the spectrum ($A_{3;\mu,00}
$) remains massless as expected by the symmetry breaking pattern. On the
other hand $A_{+;\mu,00}$ acquires a mass through the Hosotani mechanism.
It is worth noting that, except for $A_{3;\mu,00}$, all the 4D vector
bosons can be expressed in terms of complex fields.

In the scalar sector, there is a mixing between $A_4$ and $A_5$. The mass
matrices are given by: \beqn \left\{
\begin{aligned}
&
\begin{pmatrix}
A^*_{3;4,nm} & A^*_{3;5,nm}
\end{pmatrix}
\left[\frac{1}{R^2}
\begin{pmatrix}
m^2 & - rnm \\
-rnm & r^2 n^2
\end{pmatrix}
\right]
\begin{pmatrix}
A_{3;4,nm} \\
A_{3;5,nm}
\end{pmatrix}\\
&
\begin{pmatrix}
A^*_{+;4,nm} & A^*_{+;5,nm}
\end{pmatrix}
\left[\frac{1}{R^2}
\begin{pmatrix}
\left(m+\frac{\theta_5}{2\pi}\right)^2 & -
r\left(n+\frac{\theta_4}{2\pi}\right)\left(m+\frac{\theta_5}{2\pi}\right)  \\
-r\left(n+\frac{\theta_4}{2\pi}\right)\left(m+\frac{\theta_5}{2\pi}\right)
& r^2 \left(n+\frac{\theta_4}{2\pi}\right)^2
\end{pmatrix}
\right]
\begin{pmatrix}
A_{+;4,nm} \\
A_{+;5,nm}
\end{pmatrix}
\end{aligned}
\right.\nn
\eeqn The mass eigenstates $g_{3;nm}
$ and $g_{+;nm}$ are massless, while $h_{3;nm}$ and $h_{+;nm}$ have masses
$M_{3;nm}$ and $M_{+;nm}$. They are given by: \beqn \left\{
\begin{aligned}
&g_{3;nm}=\frac{m A_{3;5,nm}+r n A_{3;4,nm}}{\sqrt{m^2+r^2 n^2}}\\
&h_{3;nm}=\frac{m A_{3;4,nm}-r n A_{3;5,nm}}{\sqrt{m^2+r^2 n^2}}\\
&g_{+;nm}=\frac{\left(m+\frac{\theta_5}{2\pi}\right)A_{+;5,nm}+r\left(n+\frac{\theta_4}{2\pi}\right)A_{+;4,nm}}{\sqrt{\left(m+\frac{\theta_5}{2\pi}\right)^2+r^2\left(n+\frac{\theta_4}{2\pi}\right)^2}}\\
&h_{+;nm}=\frac{\left(m+\frac{\theta_5}{2\pi}\right)A_{+;4,nm}-r\left(n+\frac{\theta_4}{2\pi}\right)A_{+;5,nm}}{\sqrt{\left(m+\frac{\theta_5}{2\pi}\right)^2+r^2\left(n+\frac{\theta_4}{2\pi}\right)^2}}.\\
\end{aligned}
\right.\nn
\eeqn If we perform a rotation toward the mass eigenbasis in
(\ref{LgaugeEff}), we find that $g$ scalar bosons play the role of
goldstone bosons. They are eaten by the vector bosons which acquire masses.
The only physical goldstone boson is $g_{3;00}$. Actually, $h_{3;00}$ is
massless too and the effective theory contains two massless scalar degrees
of freedom, what could be a drawback.

We can now find all the interaction terms in the right basis. In addition
to the fermion-fermion-vector and fermion-fermion-scalar interactions, we
have still a bunch of vector-scalar interactions implying 3 or 4
particles. We will not write all of them but focus ourselves on the one
participating in the one loop diagrams for the EDM. These are the 3
particles interactions with at least one $A_{3;\mu,00}$ boson (the
external "photon"). They come from the following part of the 6D
lagrangian: \beqn \mathcal{L}&\supset & ie \left[A_+^{\nu}
A_+^{*\mu}\partial_{\nu} A_{3;\mu} -A_{+;4} A_+^{*\mu}\partial_{4}
A_{3;\mu} -A_{+;5} A_+^{*\mu}\partial_{5} A_{3;\mu}\right]\nn \\
&+&ie\left[ \left(\partial_{\nu}
A_{+;\mu}-\partial_{\mu}A_{+;\nu}\right)A_+^{*\nu} A_3^{\mu} -
\left(\partial_{4} A_{+;\mu}-\partial_{\mu}A_{+;4}\right)A_{+;4}^*
A_3^{\mu} - \left(\partial_{5}
A_{+;\mu}-\partial_{\mu}A_{+;5}\right)A_{+;5}^* A_3^{\mu}\right]+h.c.\nn
\eeqn Let us now introduce the mode decompositions (with only the mode
$(00)$ for $A_{3;\mu}$) to yield: \beqn \mathcal{L}&\supset & ie \
A_{+;nm}^{*\mu} A_{+;nm}^{\nu} \partial_{\nu} A_{3;\mu,00} +ie
\left(\partial_{\nu}
A_{+;\mu,nm}-\partial_{\mu}A_{+;\nu,nm}\right)A_{+;nm}^{*\nu}
A_{3;00}^{\mu}\nn \\
&+& eM_{+;nm}\ g^*_{+;nm} A_{+;\mu,nm}A_{3,00}^{\mu} +ie\
g_{+;nm}^*\partial_{\mu}g_{+;nm} A_{3;00}^{\mu} +ie\
h_{+;nm}^*\partial_{\mu}h_{+;nm} A_{3;00}^{\mu}+h.c.\nn
\eeqn

\begin{figure}[h]
\centering \includegraphics[scale=0.75]{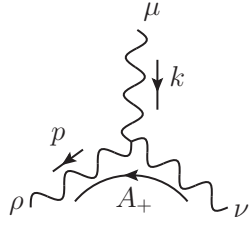}
\caption{$e
(p_{\nu}g_{\mu\rho}+p_{\rho}g_{\mu\nu}-2p_{\mu}g_{\nu\rho}+k_{\nu}g_{\mu\rho}+k_{\mu}g_{\nu\rho}-2k_{\rho}g_{\mu\nu})$.
\label{fig:diag1}
}
\end{figure}

\begin{figure}[h]
\centering
\subfloat[]{\includegraphics[scale=0.75]{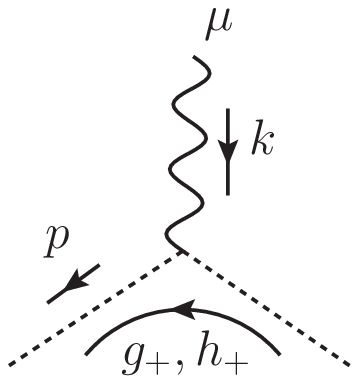}}\quad\quad\quad
\subfloat[]{\includegraphics[scale=0.75]{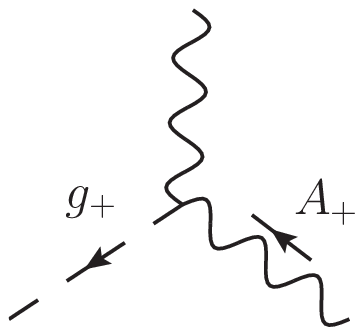}} \quad\quad\quad
\subfloat[]{\includegraphics[scale=0.75]{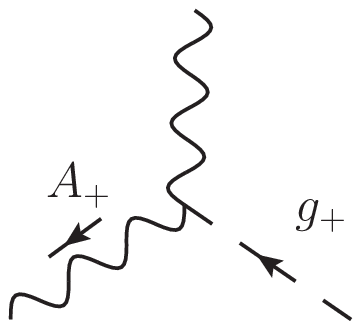}}
\caption{(a)\ $=e (2p_{\mu}-k_{\mu})$ \quad ; \quad (b) and (c)\ $=e
M_{+}$.
\label{fig:diag2to4}
}
\end{figure}

\begin{figure}[h]
\centering
\subfloat[]{\includegraphics[scale=0.75]{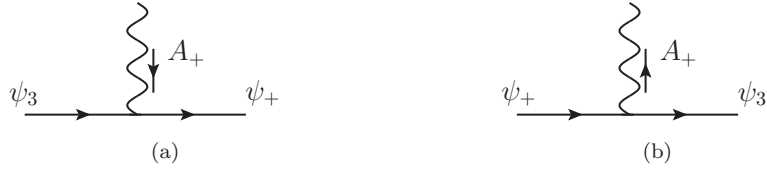}}\quad\quad\quad\quad\quad\quad
\subfloat[]{\includegraphics[scale=0.75]{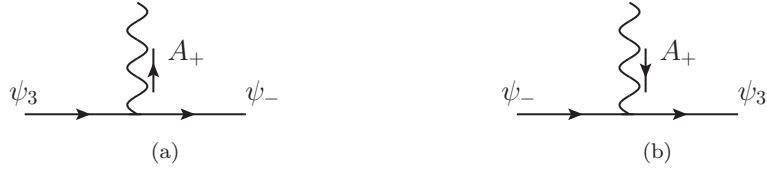}}
\caption{(a)\ $=-e
\exp\left[-i(\varphi-\varphi_+)\gamma^5/2\right]\gamma^{\mu}$\quad  ;\quad
(b)\ $=-e \exp\left[i(\varphi-\varphi_+)\gamma^5/2\right]\gamma^{\mu}$.
\label{fig:diag5to6}
}
\end{figure}

\begin{figure}[h]
\centering
\subfloat[]{\includegraphics[scale=0.75]{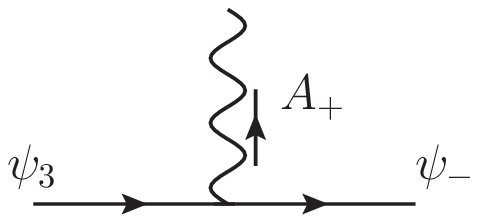}}\quad\quad\quad\quad\quad\quad
\subfloat[]{\includegraphics[scale=0.75]{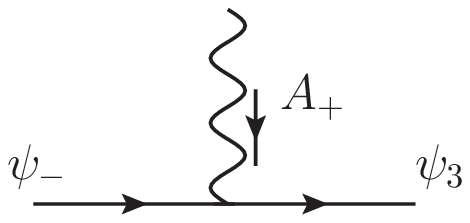}}
\caption{(a)\ $=e
\exp\left[-i(\varphi-\varphi_-)\gamma^5/2\right]\gamma^{\mu}$\quad  ;\quad
(b)\ $=e
\exp\left[i(\varphi-\varphi_-)\gamma^5/2\right]\gamma^{\mu}$.
\label{fig:diag7to8}
}
\end{figure}

\begin{figure}[h]
\centering
\subfloat[]{\includegraphics[scale=0.75]{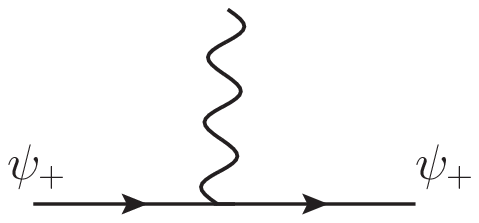}}\quad\quad
\subfloat[]{\includegraphics[scale=0.75]{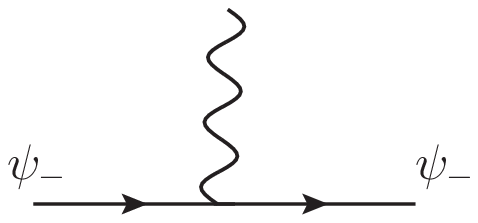}}
\caption{(a)\ $=e \gamma^{\mu}$ \quad ; \quad (b)\ $=-e \gamma^{\mu}$. \label{fig:diag9to10}}
\end{figure}

\begin{figure}[h]
\centering
\subfloat[]{\includegraphics[scale=0.7]{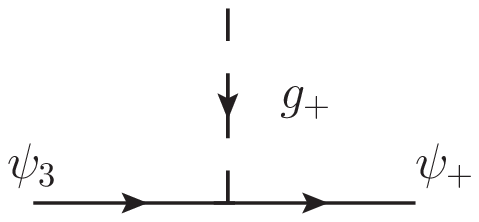}}
\subfloat[]{\includegraphics[scale=0.7]{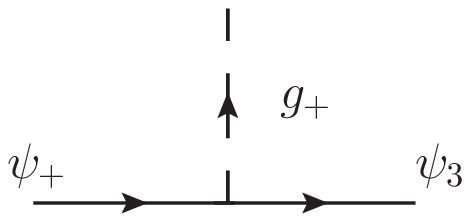}}
\subfloat[]{\includegraphics[scale=0.7]{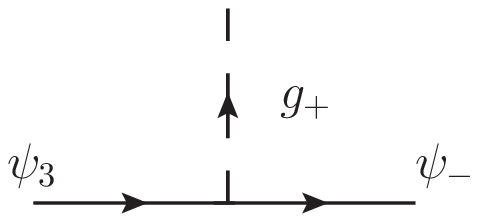}}
\subfloat[]{\includegraphics[scale=0.7]{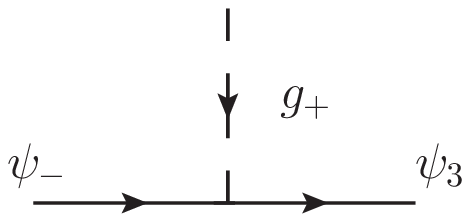}}
\caption{(a) and (b) \ $=-e \exp\left[i\left(\frac{\varphi+\varphi_+ -2\varphi^0_+}{2}\right)\gamma^5\right]$\ ; \ (c) and (d) \ $=-e \exp\left[i\left(\frac{\varphi+\varphi_- -2\varphi^0_-}{2}\right)\gamma^5\right]$.\label{fig:diag11to14}}
\end{figure}

\begin{figure}[h]
\centering
\subfloat[]{\includegraphics[scale=0.7]{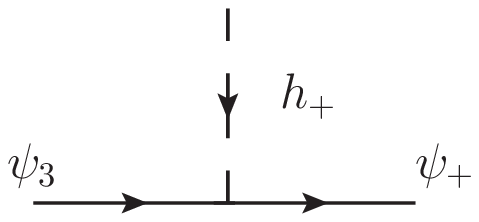}}
\subfloat[]{\includegraphics[scale=0.7]{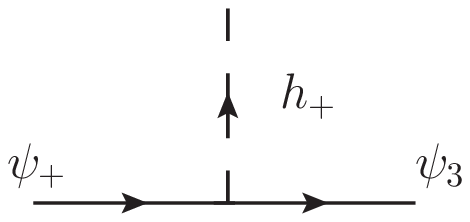}}
\subfloat[]{\includegraphics[scale=0.7]{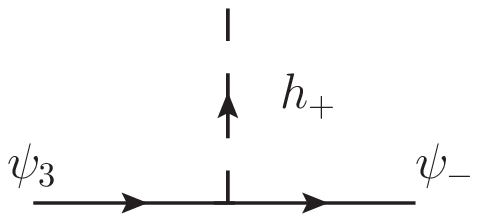}}
\subfloat[]{\includegraphics[scale=0.7]{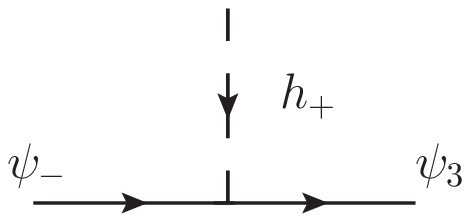}}
\caption{(a) and (b) \ $=e \exp\left[i\left(\frac{\varphi+\varphi_+ -2\varphi^0_+ +\pi}{2}\right)\gamma^5\right]$ ;  (c) and (d) \ $=e \exp\left[i\left(\frac{\varphi+\varphi_- -2\varphi^0_- +\pi}{2}\right)\gamma^5\right]$.\label{fig:diag15to18}}
\end{figure}

We give all the corresponding diagrams below (Figures \ref{fig:diag1} to
\ref{fig:diag15to18}). The additional diagrams are for the
fermion-fermion-vector or fermion-fermion-scalar interactions implying at
least one $A_{3;\mu,00}$ boson or a $\psi_{3;00}$ fermion. We use
simplified notations: $\psi_3=\psi_{3;00}$, $A_{+}=A_{+;\mu,nm}$,
$\psi_{\pm}=\psi_{\pm;\pm n \pm m}$, $g_+=g_{+;nm}$, $h_{+}=h_{+;nm}$,
$M_+=M_{+;nm}$, $\varphi=\varphi_{3;00}$, $\varphi_{\pm}=\varphi_{\pm;\pm
n\pm m}$ and $\varphi_{\pm}^0=\varphi_{\pm;\pm n\pm
m}(\beta_4=\beta_5=0)$. Finally, all the wiggled lines without label are
$A_{3;\mu;00}$ "photons". Note that charge conservation combined with
momentum conservation in ED imposes $A_{+;nm}$ interacts with
$\psi_{3;00}$ and either $\psi_{+;nm}$ or $\psi_{-;-n-m}$.

\section{Electric dipole moment of $\psi_{3;00}$}\label{AppB}

\begin{figure}[h]
\centering
\subfloat[]{\includegraphics[scale=0.75]{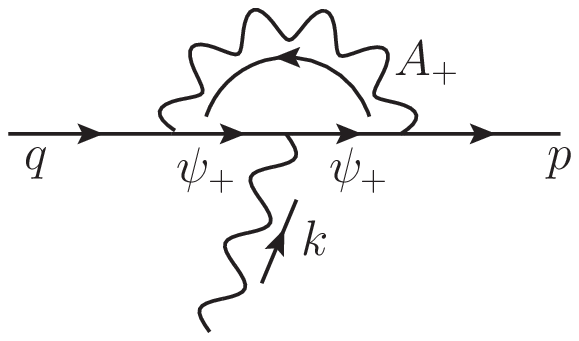}}
\subfloat[]{\includegraphics[scale=0.75]{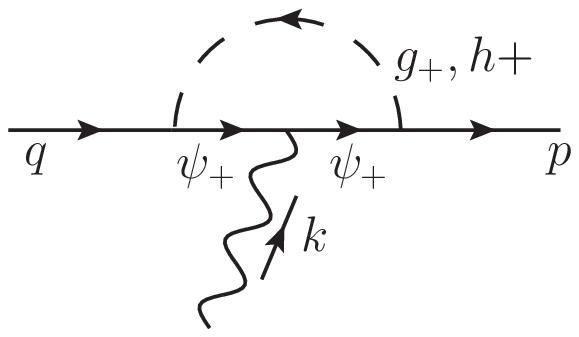}}
\subfloat[]{\includegraphics[scale=0.75]{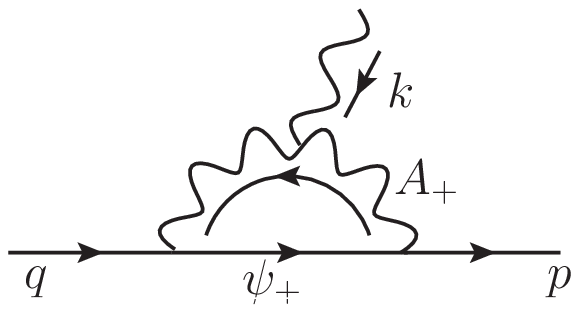}}\\
\subfloat[]{\includegraphics[scale=0.75]{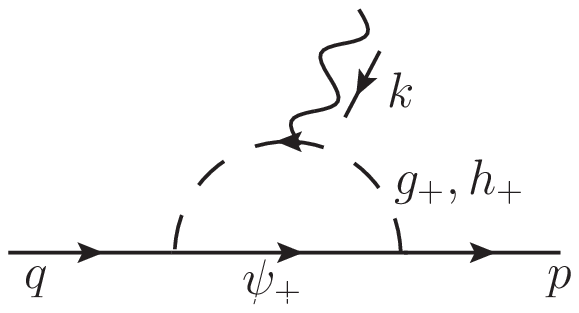}}
\subfloat[]{\includegraphics[scale=0.75]{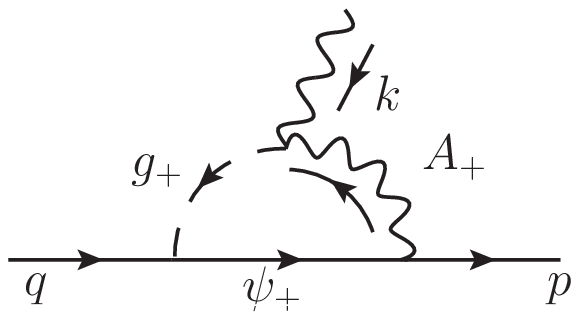}}
\subfloat[]{\includegraphics[scale=0.75]{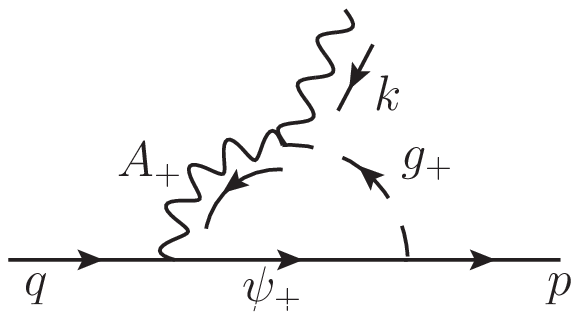}}
\caption{One-loop contributions to EDM for $\psi_{3;00}$. Contributions with $\psi_{-;-n-m}$ in the loop must be included as well. \label{fig:EDMdiag}}
\end{figure}

Six kinds of diagrams are involved in the one loop evaluation of the EDM
for $\psi_{3;00}$. We show them for a $\psi_{+;nm}$ in the loop in Figure
\ref{fig:EDMdiag}. For the $\psi_{-;-n-m}$, the fields $A_+$, $g_+$ and
$h_+$ must be replaced by complex conjugate fields (or the arrows
reversed). At the end we must sum up the $+$ and $-$ contributions and sum
over all $nm$ modes.

The contributions to $F_{nm}$ come from diagrams \ref{fig:EDMdiag}a and
\ref{fig:EDMdiag}c. The diagrams \ref{fig:EDMdiag}b and \ref{fig:EDMdiag}d
give no contributions. Finally the contributions to $J_{nm}$ and $K_{nm}$
come from diagrams \ref{fig:EDMdiag}e and \ref{fig:EDMdiag}f.

\beqn
F_{nm}^{\pm}&=&\pm\frac{\hat{m}_{\pm;\pm n\pm m}}{(4\pi)^2}
\left[4\int_0^1 \mathrm{d}x\ \frac{x(1-x)}{\Delta_{\pm;nm}(x)} +3 \int_0^1
\mathrm{d}x\ \frac{(1-x)^2}{\tilde{\Delta}_{\pm;nm}(x)}\right] \nn \\
J_{nm}^{\pm}&=&\pm\frac{R \cos \varphi_{\pm}^0}{(4\pi)^2} \int_0^1
\mathrm{d}x\ \frac{(1-x)}{\tilde{\Delta}_{\pm;nm}(x)} \nn \\
K_{nm}^{\pm}&=&\mp\frac{R \sin \varphi_{\pm}^0}{(4\pi)^2} \int_0^1
\mathrm{d}x\ \frac{(1-x)}{\tilde{\Delta}_{\pm;nm}(x)}, \nn
\eeqn where the functions $\Delta_{\pm;nm}
(x)$ and $\tilde{\Delta}_{\pm;nm}(x)$ are polynomials given by: \beqn
\Delta_{\pm;nm}(x)&=& \hat{m}^2_{3;00}\
x^2+(\hat{M}^2_{+;nm}-\hat{m}^2_{\pm;\pm n\pm m}-\hat{m}^2_{3;00})\ x+
\hat{m}^2_{\pm;\pm n\pm m}\nn \\
\tilde{\Delta}_{\pm;nm}(x) &=& \hat{m}^2_{3;00}\
x^2-(\hat{M}^2_{+;nm}-\hat{m}^2_{\pm;\pm n\pm m}+\hat{m}^2_{3;00})\ x+
\hat{M}^2_{+;nm}, \nn
\eeqn and all the "hat masses" ($\hat{m}
$, ...) are the dimensionless masses $Rm$ (the $R$ factor being factorized
in the expression of $d_E$).

It is now easy to check the two first relations (\ref{relBetaEq}). To
compute $F^{\pm}_{mn}$, we must exchange $n$ and $m$ in all the masses
$m_{\pm; \pm n \pm m}$ and $M_{+;nm}$. Since $\theta_4=\theta_5$ (imposed
by the effective potential), and $\beta_4=\beta_5$ (imposed by hand as an
hypothesis), these stay unchanged and $F$ as well. In the same way
$J_{nm}^{\pm}$ and $K_{nm}^{\pm}$ do not change through
$\tilde{\Delta}_{\pm;nm}$ but only through the phase $\varphi^0_{\pm}$.
The expressions (\ref{Cphases}) with $\theta_4=\theta_5$
($\beta_4=\beta_5=0$ by definition) show that exchanging $n$ and $m$ is
equivalent to exchanging real and imaginary part, or $\cos
\varphi^0_{\pm}$ and $\sin \varphi^0_{\pm}$. We then conclude easily that
$J^{\pm}_{mn}=-K^{\pm}_{nm}$.

\section{$+$ and $-$ chiralities sector}\label{AppC}

The only difference between $+$ and $-$ chirality lagrangians is the
matrices used to form a 6D vectors that can be "contracted" with covariant
derivative. For $+$ these are $\bar{\Sigma}^A$ matrices defined in
Appendix \ref{AppA}, and for $-$ these are $\Sigma^A$ matrices, defined as
$\Sigma^0=\bar{\Sigma}^0$ and $\Sigma^{A\neq 0}=-\bar{\Sigma}^{A\neq 0}$.
At the 4D level, an other difference appears because $\Psi_+ \sim
\begin{pmatrix} \psi_R \\ \psi_L \end{pmatrix}$, while $\Psi_- \sim
\begin{pmatrix} \psi_L \\ \psi_R \end{pmatrix}$. To form the usual Dirac
kinetic terms, we have to rewrite the lagrangians in terms of the matrices
$\bar{\Sigma}^A \gamma^0=(\gamma^{\mu},i\gamma^5,-1)$ and $\gamma^0
\Sigma^A=(\gamma^{\mu},i\gamma^5,1)$. Thus, the only remaining difference,
is a sign in the fifth component of the covariant derivative $D_5$. This
doesn't change the mass spectrum, but only the chiral phases and the
interactions with $A_5$ bosons (see Appendix \ref{AppA} for more details).
This is equivalent to change sign of $m$, $\beta_5$ and $\theta_5$ in
(\ref{Cphases}), sign of $\varphi_{\pm}^0$ and $e$ in diagrams of Figure
\ref{fig:diag11to14} (interaction with $g_+$ bosons) and sign of
$\varphi_{\pm}^0$ in diagrams of Figure \ref{fig:diag15to18} (interaction
with $h_+$ boson). Now we see that $F^{\pm}_{nm}$ and $J^{\pm}_{nm}$ don't
change, while $K^{\pm}_{nm}$ changes sign. But we must not forget that
$J^{\pm}_{nm}$ and $K^{\pm}_{nm}$ come from interaction with one $g_+$
boson (see Appendix \ref{AppB}), then they undergo an additional change of
sign. We have then the following transformations: \beqn
F^{\pm}_{nm}\sin(\varphi_{3;00}-\varphi_{\pm;\pm n\pm m}) &\longrightarrow
& F^{\pm}_{nm}\left(-\sin(\varphi_{3;00}-\varphi_{\pm;\pm n\pm m})\right)
\nn \\
J^{\pm}_{nm}\sin \varphi_{3;00} & \longrightarrow & \left(-J^{\pm}_{nm}\right)\sin \varphi_{3;00} \nn \\
K^{\pm}_{nm}\cos \varphi_{3;00} & \longrightarrow & K^{\pm}_{nm}\left(-\cos \varphi_{3;00}\right), \nn
\eeqn
which lead to the conclusion that $d_E$ changes sign (see (\ref{EDMana})).

Therefore, if we choose the same SchSch phases for $+$ and $-$ chiralities, we end up with two fermionic sectors (let us call them $P$- and $M$-sectors, which interact only through gauge and scalar interactions) with exactly the same mass spectra, and with an equal and opposite EDM for the two lightest modes.

\end{document}